\documentclass[conference]{IEEEtran}
\IEEEoverridecommandlockouts
\usepackage{cite}
\usepackage{amsmath,amssymb,amsfonts}
\usepackage{algorithmic}
\usepackage{graphicx}
\usepackage{textcomp}
\usepackage{xcolor}
\usepackage{svg}
\usepackage{amsmath}

\usepackage{subcaption}
\usepackage{mwe} 

\graphicspath{ {./img} }
\def\BibTeX{{\rm B\kern-.05em{\sc i\kern-.025em b}\kern-.08em
    T\kern-.1667em\lower.7ex\hbox{E}\kern-.125emX}}
\begin{document}

\title{A Comparison of Interfaces for Learning How to Play a Mixed Reality Handpan \\}



\author{\IEEEauthorblockN{Gavin Gosling, Ivan-teofil Catovic, Ghazal Bangash,  Daniel MacCormick, Loutfouz Zaman}
\IEEEauthorblockA{\textit{Faculty of Business and Information Technology} \\
\textit{Ontario Tech University}\\
Oshawa, Ontario, Canada \\
loutfouz.zaman@ontariotechu.ca}

}

\maketitle

\begin{abstract}

In the realm of music therapy, Virtual Reality (VR) has a long-standing history of enriching human experiences through immersive applications, spanning entertainment games, serious games, and professional training in various fields. However, the untapped potential lies in using VR games to support mindfulness through music. We present a new approach utilizing a virtual environment to facilitate learning how to play the \textit{handpan} – an instrument in the shape of a spherical dish with harmonically tuned notes used commonly in the sound healing practice of mindfulness.  In a preliminary study we compared six interfaces, where the highlighted path interface performed best. However, participants expressed preference for the standard interface inspired by rhythm games like \textit{Guitar Hero}.
 
\end{abstract}


\section{Introduction}

Virtual Reality (VR) has expanded its applications beyond entertainment, now encompassing areas such as healthcare, education, and notably, music education. It offers an innovative platform for teaching instruments, evaluating methodologies, and therapeutic interventions \cite{hobs8}. Furthermore, VR's incorporation in music has furthered its utility in healthcare, opening avenues for unique musical engagements and interventions in pain management, rehabilitation \cite{corea}, relaxation \cite{8686951}, and anxiety reduction \cite{Chirico2020}.

VR is revolutionizing accessibility in both music education and mindfulness. It provides immersive experiences, enabling users to interact with virtual instruments and partake in virtual ensembles, making learning more intuitive and inclusive. Scholars have highlighted VR's potential as an influential tool in designing therapeutic games for music therapy \cite{mti5030011},\cite{Byrns2020},\cite{pedersen:hal-03817710}.

Research indicates VR's effectiveness in music education for varied audiences. Costa et al. \cite{paper123} introduced \emph{Songverse}, a VR-based Digital Musical Instrument (DMI), facilitating intuitive music creation. Meanwhile, Shahab et al. \cite{Shahab2021} investigated the feasibility of virtual music education for children with autism, aiming to improve their social skills through realistic VR simulations. The outcomes suggested a consistent improvement in participants' musical capabilities during the sessions.

Unlike entertainment-focused video games, \emph{serious games}, designed to educate users, often struggle with player engagement \cite{7757401}. When applied to the creation of casual games, game design concepts have the potential to improve the engagement of serious games. Therefore, the use of a design approach is very important when creating VR games based on mindfulness. First of all, it is important to use an iterative and user-centered approach while creating VR solutions. The design process enables the discovery and resolution of potential barriers before formal review by addressing low-level design difficulties in the early phases. Individuals can then focus on the fundamental ideas related to the application, improving their overall experience. Second, the design procedure can help emphasize the significance of taking into account more fundamental elements that assist mindfulness practise \cite{Kelly2022}. Studies have uncovered that the combination of mindfulness and music yields a range of benefits, including heightened musical experiences, improved effectiveness of music therapy approaches such as Guided Imagery and Music (GIM), and positive contributions to mental well-being by reducing stress, providing emotional support, and fostering self-awareness. The integration of mindfulness and music not only enhances the enjoyment and depth of musical engagement but also amplifies the therapeutic effects and promotes overall mental wellness through various means\cite{hwang2021integrative}.

In our work, we develop and explore the benefits of a mixed reality (MR) rhythm video game that combines the art of playing the \emph{handpan}, a unique musical instrument, with the principles of music therapy to cultivate mindfulness skills. This simulation involves the presence of a virtual handpan in the environment with its orientation in coordination with a real handpan used by the player. The orientation is tracked by placing an orientation tracker inside the real handpan. The simulation guides the user to strike the right dimple of the handpan at each time, which  helps the user practice and learn using the handpan without the help of anyone. 

\section{Related Work}
The integration of VR technology with music therapy has emerged as a promising approach that combines the immersive power of VR with the therapeutic benefits of music. Recently, the interest to explore the potential synergies between these two domains to enhance the effectiveness of therapy and provide immersive and engaging experiences for individuals seeking therapeutic interventions, has grown.

\subsubsection{Affective computing and interfaces for mindfulness} Roo et al. \cite{Roo} designed an environment to support mindfulness practices, which features an augmented sandbox. Users can shape the sand to create a 'breathing' mini-world, which is projected back onto the sand. The user's focus is maintained through the garden's natural components, which are synced with live readings of the user's breathing patterns. Using a VR headset, users can navigate the garden for a session of meditation. Gu and Frasson \cite{Gu} presented a VR neuro-feedback relaxation training system. This system prompts users to heed vocal sophrology instructions while providing real-time feedback derived from the Meditation Score gathered by EEG. An evaluation revealed that the system decreases anxiety and depression and increases meditation scores. Frey et al. \cite{Frey2018BreezeSB} developed visual, audio and haptic output modalities for breathing and created a wearable pendant which measures breathing and transmits biofeedback in real time. They found that users intentionally modify their own breathing to match the biofeedback.

Numerous studies have focused on various mindfulness techniques as effective means of addressing anxiety and stress, utilizing approaches such as \emph{hatha yoga} and \emph{body scan} \cite{Call2013}. Building upon this research, Tang et al. \cite{Tang2020} conducted a noteworthy investigation to explore the extent to which personality traits and dispositional markers can predict individual preferences for four prototypical mindfulness techniques within mindfulness-based interventions (MBIs). Moreover, their study sought to ascertain whether these preferences hold meaningful implications for individuals' ability to achieve a meditative state. Notably, the results revealed that preferences for two specific techniques, namely loving-kindness and open monitoring, were significantly associated with related personality traits. Open monitoring involves maintaining mindfulness while allowing various stimuli, such as visualization, sound, and smell, to flow through one's awareness.

When exploring stimulation in the context of mindfulness, Graham \cite{Graham2010ACP} delves into the potential advantages of mindful music listening. Graham suggests that using music or sound as a focal point allows individuals to redirect their attention between external stimuli and internal thoughts and emotions. Through focused listening exercises, Graham proposes that this process of cultivating attentional capacity can enhance awareness and increase tolerance towards unproductive and repetitive thoughts.

\subsubsection{AR interfaces for learning music and art} Margoudi et al. \cite{maria} reviewed game-based music methods and instruments and found that despite the availability of immersive technologies few attempts have been made to leverage them for enhancing the experience of musicians. Margoudi et al. came up with a list of recommendations on how this can be improved, which include more feedback systems, less competition, incorporation of wide variety of skills in defining success, accommodation of current practices and tools of musicians, and collaboration. Löchtefeld et al. \cite{chtefeld} presented an AR system designed to help guitar students mastering their skills. This system employs a mobile projector, attached to the guitar's headstock, which casts instructional cues onto the strings. Similar approach has been used for the piano \cite{Raymaekers}, Koto \cite{Doi2017KotoLS}, Guqin \cite{Zhanginproceedings}, culinary arts \cite{Uriu} and using LEDs \cite{Seol2016LearningGW}. Fukushima and Naemura \cite{Fukushima2015WobbleSS} presented a system that facilitates the real-time observation of the rolling shutter effect through the use of  spatially divided stroboscopic projection. By animating the sweep lines, this system creates a fluctuating slow motion effect and modifies the color and texture of the guitar strings via the projected color and texture sweep lines. Shin et al. \cite{Shininproceedings} used sensors to detect the fretting hand's finger positions, which were then used  to detect the strings that are picked to measure the guitar player’s performance.

\subsubsection{VR interfaces for learning music and art} Numerous studies have examined the influence of VR and music on mental and emotional well-being \cite{Carnovalini2022} \cite{Alexanian2022} \cite{Partesotti2018}. A study by Seabrook et al. \cite{seabrook} indicates that a well-designed VR application has the potential to bolster mindfulness practice through amplifying state mindfulness and evoking positive emotions. VR technology can effectively tackle the challenges associated with mindfulness practice by creating a sensation of presence within a customized virtual environment. It allows users to engage with visual and auditory stimuli that they personally resonate with, thereby facilitating attentional focus. Additionally, VR reduces the likelihood of mind wandering by offering tailored content that aligns with the individual's mindfulness goals. Through the integration of guided mindfulness exercises and personalized virtual environments, VR uniquely enables individuals to cultivate an enhanced sense of present-moment awareness. Previously, different kinds of VR-powered meditation systems were introduced. Incorporating a head-mounted display Kosunen et al. \cite{Kosunen2016} have created a  neuroadaptive VR meditation system with the help of which individuals can float within a virtual world through meditation and mindfulness exercises. This system utilizes EEG technology to monitor users' brain activity in real time, providing estimates of their concentration and relaxation levels. Although they shifted their focus from realism and instead used minimalistic visual language in the VR scenery.
Numerous studies have explored the integration of VR and mindfulness in the treatment of various disorders. For instance, Gomez et al. \cite{Gomez2017} conducted a study utilizing Immersive VR to enhance mindfulness skills training within Dialectical Behavioral Therapy. Their work focused on a patient with severe burn injuries, aiming to lessen adverse emotions and enhance positive ones. In the immersive computer simulation, the patient engaged in mindfulness training instructions accompanied by the soothing sounds of birds chirping and flowing water, simulating the experience of floating downstream. This innovative approach highlights the potential of combining VR and mindfulness to improve emotional well-being and support individuals in therapeutic contexts.

The integration of VR, music therapy, and mindfulness has been the subject of several studies, highlighting the positive effects of incorporating soft background music into the VR environment  \cite{MA2023310} \cite{Suvajdzic2018}. Additionally, researchers have developed system frameworks that merge expressive art therapy and music therapy within VR spaces, utilizing real-time monitoring of EEG, heart rate, and gestures of participants \cite{song}. However, it is noteworthy that no research to date has explored the use of a real musical instrument in conjunction with VR for mindfulness practice.

\section{Method}

\begin{figure}[t]
\centering
\includegraphics[width=0.6\linewidth]{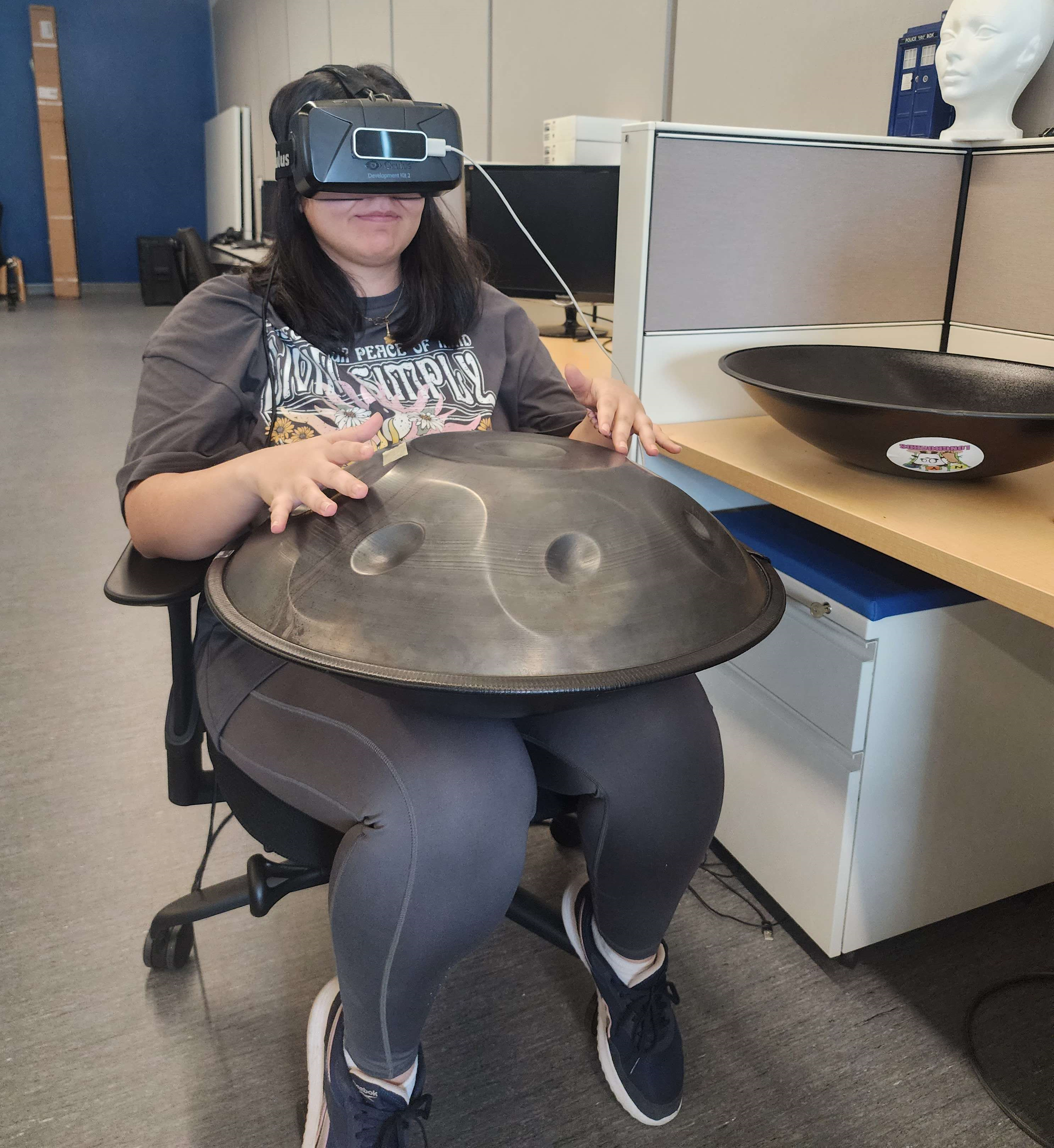}
\caption{A player interacting with the mixed-reality handpan.}
\label{fig:userimage}
\end{figure}

Our objective is to create an environment which assists the player in learning and practicing how to use the handpan. 

\subsection{The MR Handpan Environment}

\subsubsection{Handpan}
The instrument that we used was an eight-note handpan by \textit{Harmonic Arts}. See Figure~\ref{fig:userimage}. This handpan was tuned in the D Integral scale: (D3), A3, Bb3, C4, D4, E4, F4, A4.

\subsubsection{Orientation tracker}

\begin{figure}[b]
\centering
\includegraphics[scale=0.5]{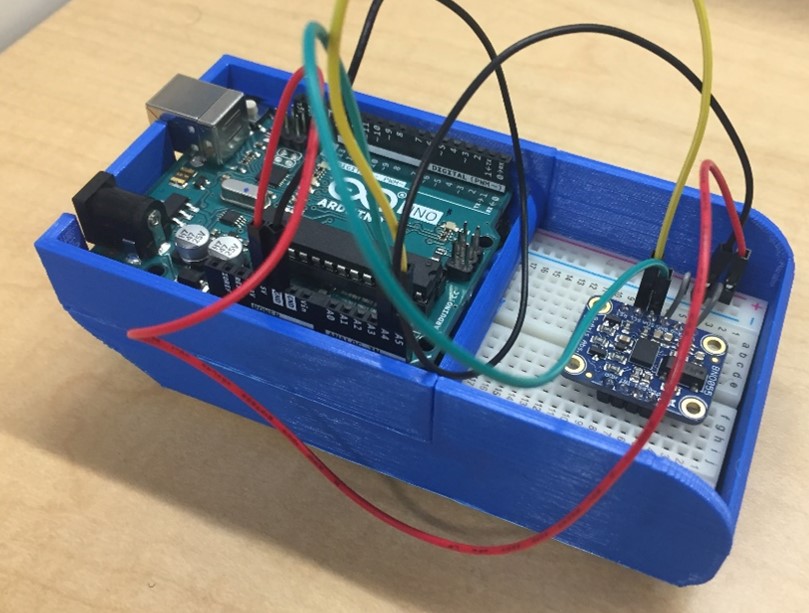}
\caption{The orientation tracking device.}
\label{fig:orientation_tracker}
\end{figure}

 The device was constructed to allow for the collection of data on the position and movement of the handpan during use. See Figure \ref{fig:orientation_tracker}. The device was programmed using the \textit{Arduino} Integrated Development Environment (IDE) and utilized appropriate libraries to access its sensors.

The device is placed inside the handpan and connected to a PC using a USB cable. The device transmits data on the orientation of the handpan to a computer. The data is collected and analyzed using the \textit{Unity} software platform, which allows for the visualization of the handpan's movements in 3D.

The primary objective of this device is to precisely track the orientation of the handpan instrument, allowing real-time replication of its motions within \textit{Unity} using game objects. The device is equipped with a comprehensive BNO 055 sensor, incorporating a gyroscope, magnetometer, and accelerometer to provide orientation and acceleration values in the form of vectors. These essential vectors are accessed by \textit{Unity} through an \textit{Arduino} sketch programmed in the \textit{Arduino} IDE, ensuring seamless communication between the physical handpan and its virtual representation in \textit{Unity}.

After soldering the BNO 055 sensor to its designated pins and establishing the correct wiring configuration with the \textit{Arduino Uno}, a script was written to access the orientation values, including a reset function and calibration checks. This script facilitated the transfer of crucial vectors to \textit{Unity}. Opening the Serial Monitor in the \textit{Arduino} IDE allowed for real-time monitoring of the collected data from the device, transmitted at a rate of 9600 bits per second.

Subsequently, the design process moved to \textit{Unity}, where a script was developed to instantiate the communication port and match the baud rate with the values in the \textit{Arduino} sketch. In the game environment, this script was attached to an empty game object called the Orientation Manager. The Orientation Manager was assigned a defined target, such as the handpan game object, enabling the device to actively change the orientation of the object within the game scene. Through careful integration, the device effectively translates the physical movements of the handpan into an interactive and synchronized virtual experience within \textit{Unity}.

By tracking the orientation of the handpan, the device provides valuable information that can be used to improve handpan playing techniques and optimize the design of future handpans. Additionally, the device can be used to develop customized training programs that are tailored to an individual's unique playing style and movement patterns. Overall, the development of this specialized tracking device represents a notable advancement in  MR handpan technology and may have the potential to enhance the playing experience for musicians.

\subsubsection{Output Device}
To enable interaction with the virtual handpan, an Oculus \textit{Rift DK2} headset was used as the output device. The Oculus \textit{Rift}  provides a fully immersive VR experience, allowing users to feel as though they are interacting with the virtual handpan in a realistic and intuitive manner.

\subsubsection{Hand Tracker}
The \textit{The Leap Motion Controller} was used to track the movement of the user's hands in real time. This allowed for a highly responsive and intuitive interaction with the virtual handpan, as the user's physical movements were accurately represented in the virtual environment. \textit{The Leap Motion Controller} is designed to track even the smallest movements of the user's hands, allowing for precise control over the virtual handpan and creating a highly realistic and engaging experience.

\subsubsection{Unity Environment}
A 3D scanner was utilized to create a detailed model of the handpan. The resulting 3D model was then imported into the \textit{Unity} software platform, where it could be manipulated and interacted with in a virtual environment. We have also developed six interfaces to interact with the handpan, described in detail below.

Overall, the combination of the 3D scanner, \textit{Unity} software, Oculus \textit{Rift} headset, and \textit{The Leap Motion Controller} allowed for the creation of a highly immersive and interactive virtual handpan experience. This technology shows promise in potentially contributing to advancements in how musicians learn and interact with handpans, offering a valuable new tool for music education and performance.

\section{Preliminary Study}

The primary objective of our preliminary study is to compare and investigate six user interfaces which we developed for learning how to play the handpan in MR. The study aims to assess the effectiveness of six interface setups in measuring participants' performance and sense of impressiveness during the learning process.

Before commencing the study, a brief tutorial was provided to familiarize participants with playing a scale using both the VR and MR handpans. During the study, participants were asked to engage with both the virtual handpan in VR  and a physical handpan in MR. They were provided with clear instructions on which dimple of the handpan to strike. Participants were presented with a song to play, and they were required to strike specific dimples on the handpan at the correct timing. They had to play two songs: one song had a total of 80 notes broken down into 10 patterns of 8, the other song had 49 notes broken down to 5 patterns of 12 (where the last pattern contained a single note). The songs were taken from \textit{Handpans and Sound Sculptures} DVD by David Kuckhermann and Colin Foulke (The corresponding videos were also taken from this DVD for the control condition with the video instruction). The order in which the interfaces were presented was counterbalanced using a 6$\times$3 Latin square.  The score was recorded based on the number of correct hits in a time span. This is an accuracy metric and it is measured in a temporal sense, meaning that the system can detect how closely the user's strike matches the intended timing of the note. Outside of the study, this also provides valuable feedback to the user, allowing them to identify areas where they need to improve their technique and timing. The study included a total of six participants which were recruited from undergraduate and graduate students in related disciplines. The experiment was a 6$\times$2 repeated measures design (6 interfaces $\times$ 2 songs). The dependent variable was score.  

By examining the effectiveness of different user interfaces in the context of handpan learning, this preliminary study aims to reveal the potential benefits and limitations of MR and VR technologies in learning how to play the handpan. Moreover, it seeks to identify the most effective interfaces for enhancing participants' learning experience and sense of impressiveness while playing the handpan.

\subsection{Interfaces}

\begin{figure*}%

    \centering
    \begin{subfigure}{0.3\textwidth}
        \includegraphics[width=\textwidth]{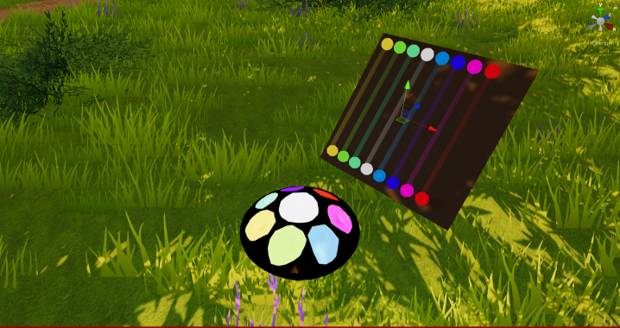}
        \caption{Standard Path}
    \end{subfigure}%
    \hfill
    \begin{subfigure}{0.3\textwidth}
        \includegraphics[width=\textwidth]{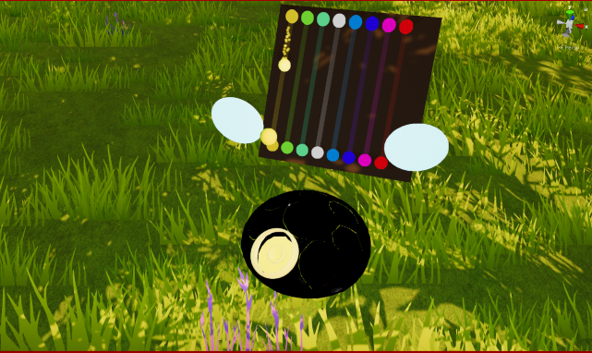}
        \caption{Highlighted Dimple}
    \end{subfigure}
    \hfill
    \begin{subfigure}{0.3\textwidth}
        \includegraphics[width=\textwidth]{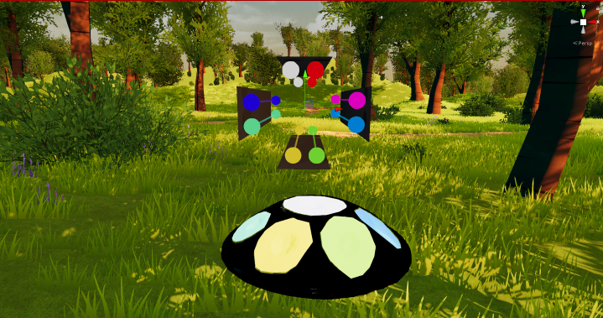}
        \caption{Four Split Path}
    \end{subfigure}

    \begin{subfigure}{0.3\textwidth}
        \includegraphics[width=\textwidth]{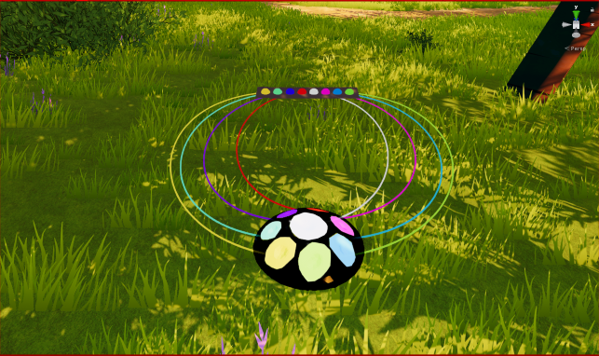}
        \caption{Direct Curved Path}
    \end{subfigure}%
    \hfill
    \begin{subfigure}{0.3\textwidth}
        \includegraphics[width=\textwidth]{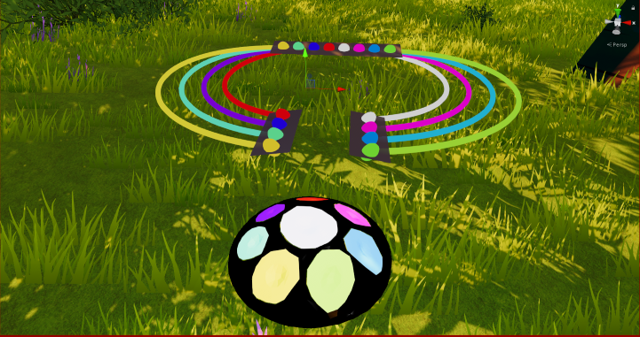}
        \caption{Semicircular Two Split Path}
    \end{subfigure}
    \hfill
    \begin{subfigure}{0.3\textwidth}
        \includegraphics[width=\textwidth]{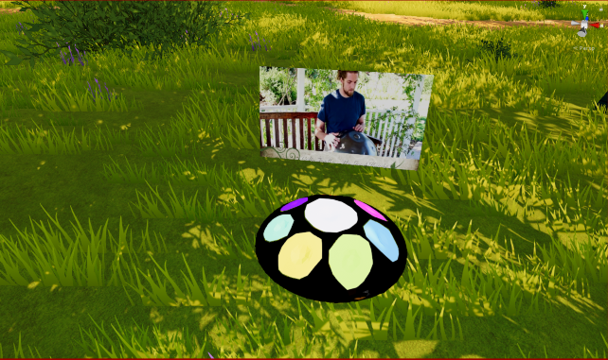}
        \caption{Video}
    \end{subfigure}
    \caption{The six interfaces.}
\label{fig:interfaces}
\end{figure*}

In all the interfaces, there are two key components: the note track and the actual notes themselves. Each of the eight tone circles on the handpan was assigned a unique color, consistently representing the notes on all interfaces. Some note tracks had distinctive interactions with the handpan game object, while others served as simple note guides. There six interfaces were as follows (see also Figure \ref{fig:interfaces}):

\subsubsection{Standard Path} This streamlined interface uses color-matching to link each line on the plane directly to a dimple on the handpan. As a droplet reaches a line's end, it signals which dimple to strike. The design draws significant inspiration from the classic rhythm game, \textit{Guitar Hero}. 

\subsubsection{Highlighted Dimple} This interface bears resemblance to the Standard Path. Furthermore, the dimple indicating the player's subsequent strike is accentuated with both highlighting and a dual-ring animation. The external ring aligns with the highlighted dimple's diameter, while the internal one begins as a central point within the dimple, expanding outward until it coincides with the outer ring. This animation aligns seamlessly with the descending note on the plane. 

\subsubsection{Four Split Path} This interface consists of four planes arranged in a tunnel formation with two notes assigned to each plane and color coded similarly to the Standard Path. 

\subsubsection{Direct Curved Path} In this interface, there is no plane of notes. Rather, the curved paths seamlessly merge into the dimples of the handpan.

\subsubsection{Semicircular Two Split Path} The paths here stem from eight distinct notes, aligning horizontally, akin to the Standard and Direct Curved pathways. At the opposing end, these notes converge into two vertical arrays.

\subsubsection{Video} This interface displays a video of a handpan player performing a song. Designed to emulate the conventional method of learning through video observation, it served as the control condition in our study.

\subsection{Results}

\begin{figure}[h]
\centering
\includegraphics[width=8cm, height=5cm]{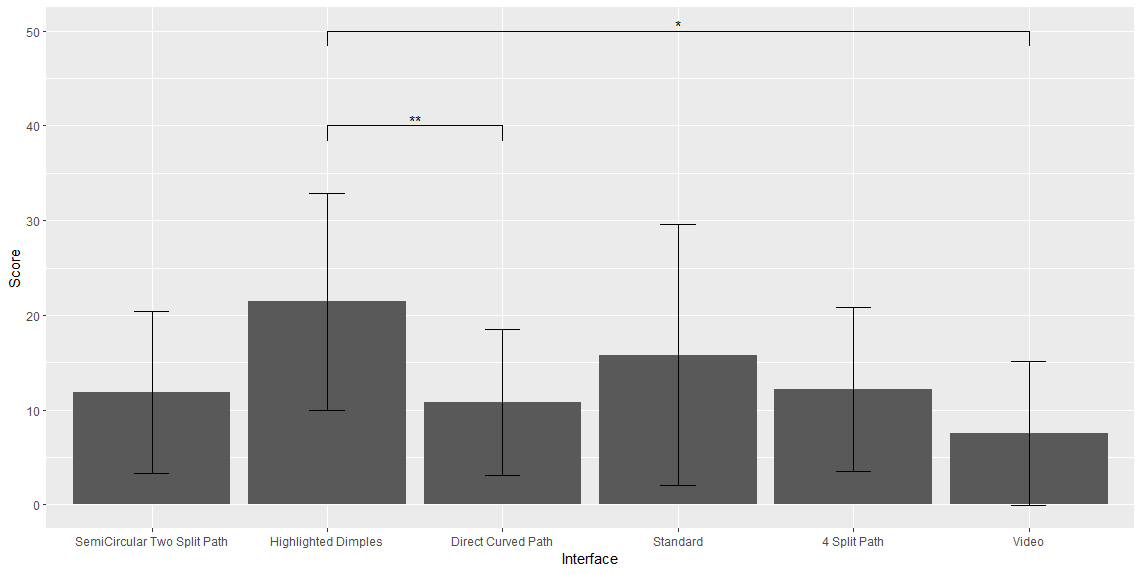}

\caption{Scores. Error bars:$\pm$1 SD, *: $p < .05$, **: $p < .01$}
\label{fig:scores}
\end{figure}

We conducted a repeated measures ANOVA test, which revealed a significant effect of Interface on Score, $\textit{F}(5,35) = 3.24, \textit{p} < .05, \eta_{G} = 0.21$. Post-hoc t-tests with Bonferroni adjustments revealed that Highlighted Dimple (\textit{M} = 21.43, \textit{SD} = 11.44) had a higher score than Direct Curved Path (\textit{M} = 10.81, \textit{SD} = 7.66), and also higher than Video (\textit{M} = 7.56, \textit{SD} = 7.58). The effect of Song was not significant, $\textit{F}(1,7) = 1.7, \textit{p} > .05$. See Figure \ref{fig:scores}.

\subsection{Participants' Feedback}

The study results revealed valuable insights from each participants' experience with the MR handpan learning system.
P1 struggled with timing and attention to the interface, primarily using one hand and facing difficulties with handpan positioning. Clear indicators for right and wrong notes were suggested, along with simplifying the complexity of interfaces, possibly introducing a speedometer, and adding labels to interfaces.
P2 faced challenges in consistently monitoring both the interface and handpan due to hardware constraints, such as latency. They recommended tweaking the GUI for scoring and addressing hand inconsistency. The Semicircular Two Split Path interface did not work well for this participant, and they expressed a lack of understanding about how to play the video.
P3 exhibited quick note responses when they lit up, but their handpan kept moving away, and yellow notes proved difficult to hit. They expressed discomfort with excessive feedback for incorrect notes but appreciated the Four Split Path interface.
P4 encountered initial difficulty in all interfaces but improved with practice, desiring slower speed and expressing dissatisfaction with the angle of instruction. They enjoyed the Standard Path interface and suggested additional visual indications.
P5 experienced confusion with unclear indications for hit or miss. The video demonstration assisted in understanding how to play the handpan, and they found the Semicircular Two Split Path interface easier for anticipating the next note. The Four Split Path interface provided too much visual data, while the Standard Path interface allowed a clearer view of the handpan. 
P6 found the handpan distracting during in the video condition, preferring highlighted notes for easy guidance. They appreciated the absence of them in the Standard Path interface but had issues with sensors and yellow notes in the Four Split Path.

Based on observations and participant feedback, the Highlighted and Standard Path interfaces were favored. Several changes were recommended for improvement, including the removal of the AUX UI, creating a tutorial for switching hands, and adding labels and descriptions to the interfaces.

\section{Conclusions and Future work}

We developed an MR handpan learning system which combines physical and virtual elements to create an interactive platform for learning and practicing handpan playing skills. 

By tracking the movements of the user's hands, orientation of the device and providing real-time feedback on their technique and timing, the environment offers a tool for musicians who are looking to improve their playing skills. The results demonstrated that the Highlighted Path interface was the most effective in terms of the measured accuracy. Subjectively, participants preferred the Standard interface, likely for being the most 'clean' and familiar one. 

The preliminary study demonstrated a potential for the environment to be effective, with participants able to seamlessly integrate the virtual and physical elements of the handpan playing experience. 

In the future, the integration of mindfulness principles into the system can enhance its potential as a transformative tool for musicians.

\section*{Acknowledgment}

We would like to extend our gratitude to Andrew Hogue for creating a 3D scan of the handpan.

\bibliographystyle{IEEEtran}
\bibliography{ref}
\end{document}